\documentclass[twocolumn,aps,prx,superscriptaddress,longbibliography]{revtex4-2}
\usepackage{graphicx}
\usepackage{color}
\usepackage{amsmath}
\usepackage[export]{adjustbox}
\usepackage{enumitem}
\usepackage{amssymb}
\usepackage{hyperref}
\usepackage[normalem]{ulem}
\usepackage{adjustbox}
\usepackage{siunitx}
\usepackage{lipsum}

\hypersetup{
	colorlinks = true,
	linkcolor = blue,
	citecolor = blue,
	urlcolor  = blue,
}

\newcommand{\be}{\begin{equation}}
	\newcommand{\ee}{\end{equation}}

\newcommand{\bea}{\begin{eqnarray}}
	\newcommand{\eea}{\end{eqnarray}}

\renewcommand{\epsilon}{\varepsilon}

\newcommand{\ket}[1]{\left|#1\right\rangle}

\DeclareUnicodeCharacter{0105}{\k{a}} 

\definecolor{mygreen}{rgb}{0.21, 0.73, 0.13}

\newcommand{\param}[2]{#1_{\mathrm{#2}}}

\begin{document}

 \title{Unified evolutionary optimization for high-fidelity spin qubit operations}

	\date{\today}
	
	\author{Sam R. Katiraee-Far}
\email{S.R.Katiraee-Far@student.tudelft.nl}
\affiliation{QuTech and Kavli Institute of Nanoscience, Delft University of Technology, 2628 CJ Delft, the Netherlands}%
\author{Yuta Matsumoto}
\affiliation{QuTech and Kavli Institute of Nanoscience, Delft University of Technology, 2628 CJ Delft, the Netherlands}
\author{Brennan Undseth}
\affiliation{QuTech and Kavli Institute of Nanoscience, Delft University of Technology, 2628 CJ Delft, the Netherlands}%
\author{Maxim De Smet}
\affiliation{QuTech and Kavli Institute of Nanoscience, Delft University of Technology, 2628 CJ Delft, the Netherlands}%
\author{Valentina Gualtieri}%
\affiliation{QuTech and Kavli Institute of Nanoscience, Delft University of Technology, 2628 CJ Delft, the Netherlands}%
\author{Christian Ventura Meinersen}
\affiliation{QuTech and Kavli Institute of Nanoscience, Delft University of Technology, 2628 CJ Delft, the Netherlands}%
\author{Irene Fernandez de Fuentes}
\affiliation{QuTech and Kavli Institute of Nanoscience, Delft University of Technology, 2628 CJ Delft, the Netherlands}%
\author{Kenji Capannelli}
\affiliation{QuTech and Kavli Institute of Nanoscience, Delft University of Technology, 2628 CJ Delft, the Netherlands}%
\author{Maximilian Rimbach-Russ}
\affiliation{QuTech and Kavli Institute of Nanoscience, Delft University of Technology, 2628 CJ Delft, the Netherlands}%
\author{Giordano Scappucci}
\affiliation{QuTech and Kavli Institute of Nanoscience, Delft University of Technology, 2628 CJ Delft, the Netherlands}%
\author{Lieven M. K. Vandersypen}%
\affiliation{QuTech and Kavli Institute of Nanoscience, Delft University of Technology, 2628 CJ Delft, the Netherlands}%
\author{Eliska Greplova}%
\email{e.greplova@tudelft.nl}
\affiliation{QuTech and Kavli Institute of Nanoscience, Delft University of Technology, 2628 CJ Delft, the Netherlands}%

\begin{abstract}
Developing optimal strategies to calibrate quantum processors for high-fidelity operation is one of the outstanding challenges in quantum computing today. 
Here, we demonstrate multiple examples of high-fidelity operations achieved using a unified global optimization-driven automated calibration routine on a six dot semiconductor quantum processor. Within the same algorithmic framework we optimize readout, shuttling and single-qubit quantum gates by tailoring task-specific cost functions and tuning parameters based on the underlying physics of each operation. Our approach reaches systematically $99\%$ readout fidelity, $>99\%$ shuttling fidelity over an effective distance of \SI{10}{\micro \meter}, and $>99.5\%$ single-qubit gate fidelity on timescales similar or shorter compared to those of expert human operators. The flexibility of our gradient-free closed loop algorithmic procedure allows for seamless application across diverse qubit functionalities while providing a systematic framework to tune-up semiconductor quantum devices and enabling interpretability of the identified optimal operation points.
\end{abstract}

\maketitle

 \section{Introduction}

Semiconductor quantum devices are one of the prominent quantum computing platforms due to their scalability through foundry-compatible fabrication methods ~\cite{zwerver2022qubits,burkard2023semiconductor,weinstein2023universal,intel_fab2024,SteinackerCMOS2024,intel_twelvespin_2025}. Rapid single-shot readout~\cite{collard_latching_2018,blumoff2022fast,takeda2024rapid} and a high-fidelity universal gate set~\cite{noiri_highfidelity_2022,xue_highfidelity_2022,mateusz_highfidelity_2022,mills_highfidelity_2022} have been demonstrated, as well as scalable two-dimensional architectures~\cite{morte_3by3_2021,federico_twod_STqubit_2021,Borsoi2024SharedControl}. Additionally, added operational flexibility of semiconductor chips has been achieved using shuttling techniques allowing for increased connectivity on the chip~\cite{yoneda2021coherent,noiri_shuttle_2022,struck_conveyor2024,geshuttling_2024,de2024high}.

Despite much recent progress in the field, bottlenecks still remain in unlocking the potential scalability of state-of-the-art technology. Generally, these boil down to variability of the devices: slight differences between individual quantum dots, unwanted background potentials, or drift of the device as a function of time. These factors yield requirements for specialized fine-tuning of individual dots during the operation, typically done by human experts. These challenges gave rise to the field of autonomous tuning and operation of semiconductor devices~\cite{van2018automated,durrer2020automated,zwolak2020autotuning,zwolak2023colloquium,zwolak2023data}. To date, this line of research has been dominated by the adaptation of computer vision techniques that autonomously reproduce human-like analysis of two-dimensional data with an emphasis on faster and less noise-sensitive performance~\cite{lennon2019efficiently,darulova2020autonomous,schuff2024fully}. While this class of methods has been immensely successful on smaller scale devices and well-specified tasks, image recognition inherently reduces each task to a low-dimensional feature identification problem that might not take into account multidimensional dependencies that may arise due to cross-talk. Furthermore, specialized single-task algorithm development makes it more challenging to re-purpose the algorithms from one task to others.

In this work, we put forward a gradient-free evolutionary algorithm approach to spin-qubit operation that can be directly applied across diverse optimization tasks with minimal modification. This method excels at high-dimensional parameter optimization, efficiently navigating the complex parameter landscapes inherent to quantum dot systems, and does not require pre-training or labeled datasets, making it immediately reusable across different experiments.

Specifically, we utilize an evolutionary algorithm as a fine-tuning agent for the automated operation of a linear array of six gate-defined quantum dots in a $\mathrm{^{28}Si/SiGe}$ heterostructure. We concentrate on key building blocks for the operation of this universal quantum dot quantum processor: readout, shuttling and gate operation. The first task that we approach is that of Pauli Spin Blockade (PSB) readout~\cite{seedhousePauliBlockadeSilicon2021}. The performance of this readout technique is known to depend on a multitude of parameters such as the pulse amplitudes and ramp times. Secondly, we aim to preserve quantum coherence of the electron wavefunction during the shuttling operation. This task can be accomplished by fine-tuning the quantum dot gate voltages to eliminate the background potentials that may interfere with the shuttling. Finally, tune-up of single-qubit gates is a ubiquitous task for quantum computing platforms. While it is less parameter intensive than readout and shuttling, there is a need for a device agnostic approach that deterministically reaches high gate fidelities. The common feature among these tasks is a potentially highly correlated parameter space intrinsic to devices with long range Coulomb interactions. To address these challenges we deploy the Covariance Matrix Adaptation Evolution Strategy (CMA-ES)~\cite{hansen2006cma,nomura2024cmaes} as it is known to adapt well to high-dimensional optimization landscapes. Moreover, the fact that this approach is gradient-free enhances its flexibility with respect to the noise. Notably, while these diverse qubit operation tasks are qualitatively very different from each other, they are optimized using a unified algorithmic framework that is straightforward to embed into any standard spin-qubit operation software.

We use the CMA-ES to maximize the PSB readout visibility, systematically reaching values $> 98\%$ corresponding to a readout fidelity of $99\%$. The reliance of the algorithm on covariances between variables also allows us to get insights into correlations between variables influencing readout quality which are not obvious from manual tuning. Secondly, we obtain $>99\%$ shuttling fidelity over an effective distance of \SI{10}{\micro \meter}. Finally, we optimize the single-qubit gate fidelity via a randomized benchmarking sequence systematically reaching values $>99.5\%$. 

\begin{figure}
    \includegraphics[width=\columnwidth]{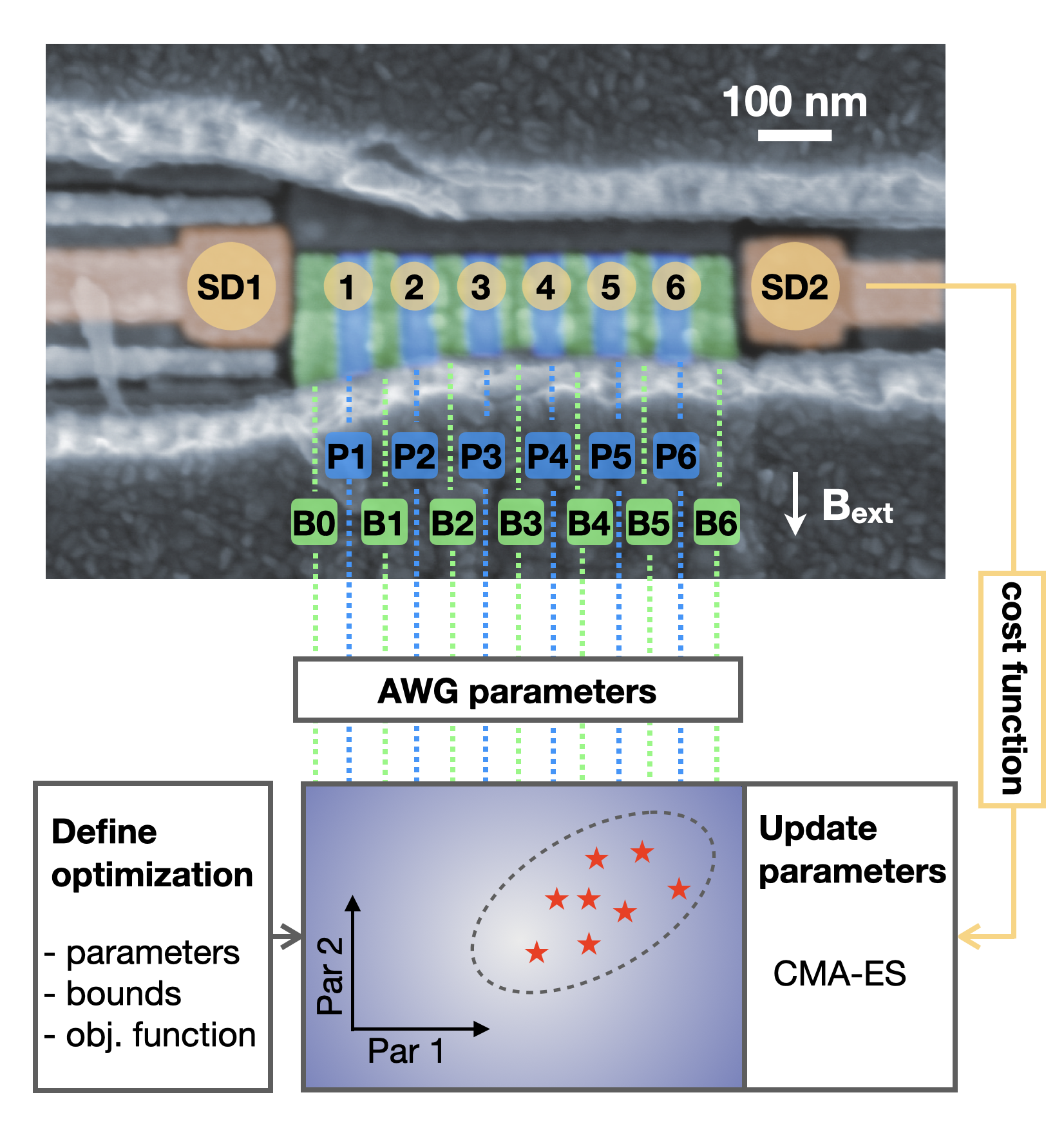}
    \caption{\textbf{Operation scheme used in the presented protocols}. The upper panel shows the linear six dot device with quantum dots denoted by number $1-6$, two sensing dots (SD$1$, SD$2$), six plunger gates in blue (P$1$ - P$6$), and seven barrier gates in green (B$0$ - B$6$). The first step in each operation protocol consists of the definition of the optimization (lower left corner): which parameters we want to adjust, within which bounds, and what quantity (cost function) we wish to use to measure the quality of operation. The readout values get combined into the cost function that expresses the quality of the current parameter setting. This value is in turn used to update the parameter distribution (schematically shown in purple in the center of the lower panel: the white region corresponds to the low cost values, purple to high cost values) via the CMA-ES. This distribution is then sampled (red stars) and the resulting values are passed back to the devices via adjustment of the AWG parameters.}
    \label{fig:scheme}
\end{figure}

\section{Results}

The common optimization framework is illustrated in Fig.~\ref{fig:scheme}. The readout outcome of the linear 6-dot device is transformed to the values of the task-specific cost function. The value of this function is used as an input for the CMA-ES~\cite{hansen2006cma,nomura2024cmaes} algorithm that updates the parameter distribution in an effort to decrease this cost function. The new parameters are sampled from the parameter distribution and uploaded to the arbitrary wave generator (AWG) controlling the device. This loop is repeated until the optimal AWG parameters for the desired task are reached. The algorithm runs autonomously, and, in the examples we report here, achieves the optimal performance at a similar pace or faster and more systematically than expert human operators. 

The device used in this work is fabricated on an isotopically-purified $^{28}$Si/SiGe heterostructure with a \SI{7}{\nano \meter} quantum well, as described in~\cite{de2024high} (see Fig.~\ref{fig:scheme}). There are two sensing dots located at the ends of the linear array to facilitate charge sensing (SD1, SD2). A cobalt micromagnet is deposited on top of the gate electrodes. When magnetized with an external magnetic field $B_{ext}$, its stray field enables the addressability of each individual electron as well as electric-dipole spin resonance (EDSR) for single-qubit rotations~\cite{obata2010coherent}. An external magnetic field of \SI{260}{\milli T} is applied in-plane with respect to the quantum well and all experiments are performed in a dilution refrigerator set to a temperature of \SI{200}{\milli K}~\cite{undseth2023hotter}.

\subsection{Readout optimization}

\begin{figure*}
    \centering
    \includegraphics[width=\linewidth]{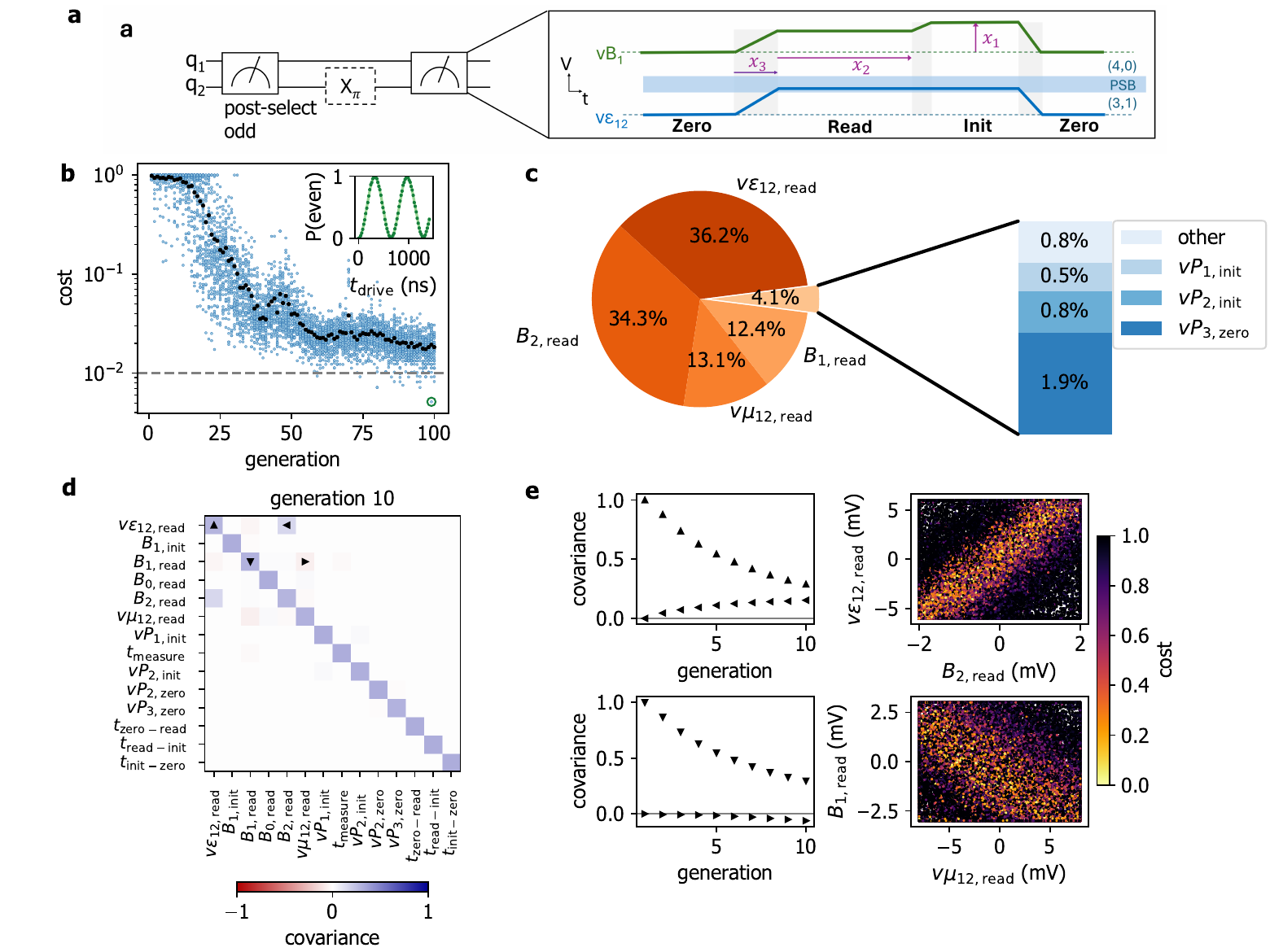}
    \caption{\textbf{Pauli Spin Blockade readout optimization and sensitivity analysis.} \textbf{a)} Two circuit variations on qubits $q_1$ and $q_2$ are used to obtain a cost function value for the PSB parity readout. In both variations, we first initialize the qubits by post-selecting the odd-parity measurement outcomes after randomly initializing the qubit pair and end the circuit by measuring the parity. Circuit variation 2 includes an additional $X_\pi$ gate on $q_2$, to prepare an even parity state. We obtain the cost function as the difference between the odd-parity fractions of the 1000 measurements between the two circuit variations (ideally 0 and 1 for the first and second variation, respectively). The PSB measurement operation is illustrated schematically by a piecewise-linear pulse, showing the amplitude as a function of time for both the detuning $v\epsilon_{12}$ and the barrier $B_{1}$. The pulse is dissected into three stages where the amplitude remains constant: ``zero", where we idle and perform gates, ``read", where the sensor signal is acquired, and ``init", where the qubits are prepared for initialization after a PSB measurement. The variables $x_i$ are shown as examples for parameters involving voltage amplitudes, durations and ramp times. \textbf{b)} The cost function values (blue dots) throughout the 14-dimensional readout optimization. The optimization was run for 100 generations with 50 individuals per generation. The black dots represent the mean cost function value per generation. The inset displays Rabi oscillations on qubit 1 after optimizing the AWG parameters, corresponding to the run marked by the green circle. \textbf{c)} A pie chart showing the parameter sensitivity analysis on 44  optimization runs, each consisting of 10 generations of 100 sampled individuals. The parameters with the highest contribution to the cost function are shown in the pie chart. The remaining parameter contributions are plotted as a bar chart. \textbf{d)}
    A heat map of the element-wise average of the 44 covariance matrices obtained with the CMA-ES at generation 10. Two diagonal elements of the matrix $v\epsilon_\mathrm{12,read}, v\epsilon_\mathrm{12,read}$ (up pointing triangle) and $B_\mathrm{1,read}, B_\mathrm{1,read}$ (down pointing triangle) and two off-diagonal elements, $v\epsilon_\mathrm{12,read},B_\mathrm{2,read}$ (left pointing triangle), and $B_\mathrm{1,read},v\mu_\mathrm{12,read}$ (right pointing triangle) are highlighted. \textbf{e)} The marked covariances plotted versus the generation. The average cost function value over the 44 optimization runs is plotted for the $v\epsilon_\mathrm{12,read},B_\mathrm{2,read}$ and $B_\mathrm{1,read},v\mu_\mathrm{12,read}$ pairs in the upper and lower panel, respectively.}
    \label{fig:psb}
\end{figure*}

For readout, we employ the (3,1)-(4,0) charge transitions of dots 1, 2, and 6, 5 respectively. We use parity Pauli Spin Blockade (PSB) \cite{seedhousePauliBlockadeSilicon2021} for spin-to-charge conversion via the pulse sequence shown in the right panel of Fig.~\ref{fig:psb}a. This pulse takes a piecewise linear form, which we dissect in three distinct stages: the readout sequence starts in stage ``zero", the measurement signal is integrated in ``read" and the spins are prepared for initialization during ``init".

The optimization variables include the amplitudes, durations, and ramp times of the different stages. We denote a voltage amplitude as $(v)\param{g}{i(j),s}$, a (virtual) gate $g$ in stage $s$, corresponding to dot $i$ (and $j$), numbered as indicated in Fig.~\ref{fig:scheme}. In total, we optimized $14$ variables. Key parameters are the voltages that control the detuning between (4,0) and (3,1), $\epsilon$, and their energy, $\mu$, during the ``read" stage, labeled as $\param{v\epsilon}{12,read}$ and $\param{v\mu}{12,read}$, respectively, as they determine the PSB readout window. The amplitude of the barrier ($B$) voltages, which control the tunnel coupling between the dots, were optimized for the ``read" and ``init" stages, specifically $\param{B}{0,read}, \param{B}{1,read}, \param{B}{2,read},$ and $\param{B}{1,init}$. Considering all three barrier gates is often necessary to properly control the tunnel coupling between dots 1 and 2.  The plunger ($P$) gate voltage amplitudes were optimized for the ``init" stage, $\param{vP}{1,init}$ and $\param{vP}{2,init}$, to compensate for the high amplitudes pulsed on $\param{B}{1,init}$, as well as the ``zero" stage, $\param{vP}{2,zero}$ and $\param{vP}{3,zero}$ to mitigate unwanted charge leakage, specific to this device. Finally, we included the measurement time $\param{t}{measure}$ and the ramp times between the three consecutive stages: $\param{t}{zero-read}$, $\param{t}{read-init}$, and $\param{t}{init-zero}$, which influence the state transfer adiabaticity.
 
We randomly sample the normalized optimization variables from a normal distribution with initial mean $\mu_i \in (0,1)$, where $i$ denotes the $i$-th optimization variable, and initial step size $\sigma = 1$. This sampling distribution is updated using the CMA-ES for $100$ generations. Each generation contains $50$ individual parameter samples, each of which we evaluate for $1000$ shots to gather readout fidelity statistics. 
Here, we use the PSB readout visibility as a cost function for the optimization. This quantity is  defined as
\begin{equation}
V = n(\mathrm{odd}|\mathrm{odd}) - n(\mathrm{odd}|\mathrm{even}), 
\label{eq:visibility}
\end{equation}

\noindent where $n(\mathrm{odd}|\mathrm{odd})$ is the fraction of odd-parity measurement outcomes, $\ket{q_1q_2} = \ket{\uparrow \downarrow} $ or $\ket{q_1q_2} = \ket{\downarrow \uparrow}$, after post-selecting on odd-parity initial states and $n(\mathrm{odd}|\mathrm{even})$ is the fraction of odd-parity measurement outcomes after preparing an even-parity state, $\ket{q_1q_2} = \ket{\uparrow \uparrow}$ or $\ket{q_1q_2} = \ket{\downarrow \downarrow}$, by virtue of applying a $X_\pi$-rotation on qubit 2. The circuit executed on the device is shown in the left panel in Fig.~\ref{fig:psb}a. We note that in order to perform this operation, prior qubit control is required. We therefore coarse tuned the device to approximately 85\% visibility prior to starting the optimization run. 

In Fig.~\ref{fig:psb}b, the evolution of the cost function~\eqref{eq:visibility} is shown against the generation of the CMA-ES run with the best value denoted by a green circle. The inset shows the Rabi oscillation obtained using the parameters extracted from the respective optimization point. By fitting the data to the function $P_{\textit{even}} = V_R \cos(\omega_{R} \param{t}{drive} + \phi ) \exp (-t/\tau)$, we obtain a Rabi visibility $V_R = 99.3 \pm 0.6 \%$.

We perform a parameter sensitivity analysis of the $14$ optimization parameters. For this, we use the data of $44$ repeated optimization runs to increase the exploration of the search space. Each run contains $10$ generations with a population size of $100$ parameter samples. 
The method we use for the sensitivity analysis is high-dimensional model representation~\cite{herman2017salib} (HDMR), which allows extracting the relative contributions of each  parameter to the chosen cost function. The most significant parameters are plotted in a pie chart in Fig.~\ref{fig:psb}c along with the contribution of several other parameters in a bar chart. Notably, the largest contributions to the readout fidelity are attributed to the plunger and barrier voltages, which, due to crosstalk effects, together determine whether the system is in the Pauli spin blockade regime. Furthermore, the initialization parameters $\param{vP}{1,init}$ and $\param{vP}{2,init}$ make particularly critical contributions to achieving $>99\%$ visibility. We analyze the importance of the initialization pulse further in Appendix~\ref{app:init-theory}. The contribution from $\param{vP}{3,zero}$ can also be attributed to either its influence on the initialization pulse or charge leakage from dot 2 to dot 3. It is important to note that while HDMR normalizes all parameters prior to the analysis, this normalization cannot take into an account the different physical meanings of the variables or the initial range we choose to explore for each variable. For instance, if we have a good empirical prior for a given variable and thus specify a narrower value range for the optimization, it will appear that this variable affects the cost function less than it would shoud we have specified a broader value range. As such, it is key to understand the percentages in Fig.~\ref{fig:psb}c as values relative to the particular setting.

The fact that the CMA-ES uses an unscaled covariance matrix (see Appendix~\ref{app:CMA-ES} for further details) to capture directional dependencies between parameters allows for a certain degree of interpretability. In Fig.~\ref{fig:psb}d, we show a snapshot of the covariance matrix at the final generation, averaged over each of the 44 repeated runs. We highlight several covariances in this covariance matrix (triangles) and track their evolution over the 10 generations of optimization in the left panel of Fig.~\ref{fig:psb}e. We observe that the variance of $\param{v\epsilon}{12,read}$ decreases, while the covariance of the pair ($\param{v\epsilon}{12,read}, \param{B}{2,read}$) increases. This positive correlation due to crosstalk is also visible in the scatter plot of the cost function values against this variable pair in the upper right panel of Fig.~\ref{fig:psb}e. Similarly, the lower two panels show the negative correlation due to crosstalk that was picked up between ($\param{v\mu}{12,read}, \param{B}{1,read}$).

To test the consistency of our approach, we deployed our algorithm to repeatedly calibrate readout on another linear 6-quantum-dot device \cite{philips2022universal} using a similar set of parameters, reaching consistent high-visibility results, see Appendix~\ref{app:readout-consistency} for details.

\subsection{Optimization of Conveyor Shuttling}
\begin{figure*}
    \centering
    \includegraphics[width=.9\linewidth]{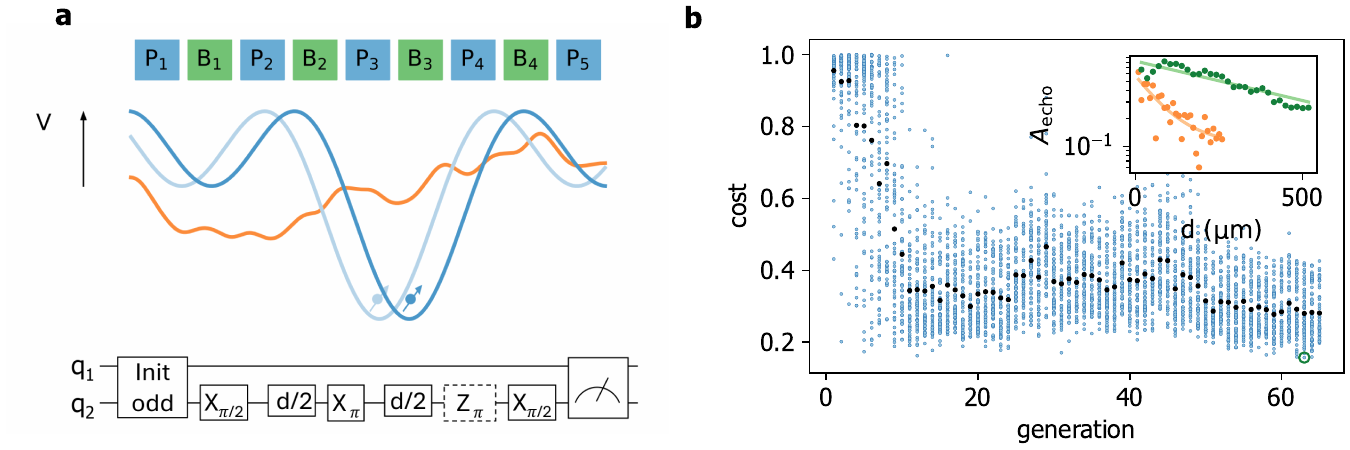}
    \caption{\textbf{Conveyor shuttling fidelity optimization.}\textbf{a)} Upper panel shows a schematic of the background potential (orange) and the applied conveyor potential (blue) at two points in time. The optimization parameters are the offset voltages on the plunger gates (blue) and barrier gates (green). Lower panel shows the circuit which is executed to obtain the echo amplitude cost function. \textbf{b)} The cost function value (blue) versus the generations of one CMA-ES run. The black dots denote the mean cost function value per generation. The inset shows how the echo amplitude, $A_{echo}$, decays with the distance $d$ prior to the optimization (orange dots) and after the optimization (green) on a log scale. Both measured datasets are fitted to the function $A = A_0 p^d + C$, to obtain $p = 0.117 \pm 0.001$ before optimization and $p =0.0192 \pm 0.0002 $ after optimization. 
    }
    \label{fig:shuttling}
\end{figure*}

\noindent Shuttling electron spins was established as a powerful tool for increasing quantum dot chip connectivity ~\cite{yoneda2021coherent,de2024high,geshuttling_2024} and constitutes a significant step forward in the development of flexible reconfigurable platforms that are readily adaptable to specific application needs. In particular, \textit{conveyor mode shuttling} has demonstrated promising high-fidelity qubit transport~\cite{de2024high}. This added flexibility is not, however, without its challenges. The shape of the moving potential in conveyor shuttling is affected by variations in the gate electrode lever arms and electrostatic crosstalk as well as disorder in the electric potential from, e.g., charge defects. This makes it challenging to sustain high fidelities during the spin transfer.

Recently, the scheme for the so-called two-tone conveyor shuttling was successfully demonstrated in Ref.~\cite{de2024high}. Here, the electron is physically transported from one end of the array to the other (specifically from under gate $P_2$ to under gate $P_5$ in Fig. 1) by a traveling wave potential produced by the combination of two sinusoidal voltages (the two-tone conveyor), see Appendix~\ref{supp:subsec:twotone}.
The visualization of this procedure as well as the corresponding gate scheme is shown in Fig.~\ref{fig:shuttling}a. While shuttling is a completely different physical process than readout, on a high level it fits within the class of protocols exemplified in Fig.~\ref{fig:scheme}: we require a real-time optimization of the set of physical parameters to optimize an objective function we can extract by measuring the device. Here, we quantify performance by comparing the state before and after shuttling. The parameters are the $8$ offset voltage pulses on the gates during shuttling, $vP_1, B_1, vP_2, B_2, vP_3, B_3, vP_4, B_4$. The goal of this optimization is to remove unwanted background potential variations (orange line in Fig.~\ref{fig:shuttling}a) that arise as a consequence of the disorder.

Since reconstructing the fidelity of the shuttled state is time-consuming, we instead use the qubit coherence as a cost function. We assume and verify that the spins are not lost by the shuttling process and we further assume that there are no residual coherent operations on the spins due to the shuttling. Thus, we attribute the loss of fidelity purely to a loss in coherence during shuttling, which we can measure using a spin echo sequence (see Fig.~\ref{fig:shuttling}a). Specifically, we use the amplitude of the measured spin-echo signal as the cost function. In experiment, the spin is shuttled back and forth for 400 rounds, covering a total distance $d$ = \SI{172.8}{\micro \meter}. The phase accumulation caused by the potential landscape experienced during shuttling and the spatially varying magnetic field due to the micromagnet is canceled with an $X_\pi$ echo pulse pulse applied on dot 2 halfway, at a distance $d/2$. We run two variants of this circuit by adding an additional $Z_\pi$-rotation on qubit 2 for one of the circuit variations. The inclusion of the additional $Z_\pi$-rotation flips the parity of the spins, thus providing contrast between the measurement outcomes of the two circuit variations.

Fig.~\ref{fig:shuttling}b shows the cost function as a function of the number of generations of the CMA-ES. We observe that the mean value of the cost converges to values $<$ 0.5 from generation 10 onward. The offset voltages corresponding to the lowest cost function value (green circle) are used to measure the spin echo amplitude as a function of a variable shuttling distance $d$ (green dots), depicted in the inset. The same measurement was also performed before running the algorithm (orange dots). Following the procedure outlined in Ref.~\cite{yoneda2021coherent}, we fit both datasets to the function $A = A_0p^d + C$, to obtain a coherence loss per 10 $\mathrm{\mu m}$ corresponding to the depolarization parameter $p = 0.117 \pm 0.001$ before optimization and $p =0.0192 \pm 0.0002$ after optimization. The \SI{10}{\micro \meter} shuttling fidelity is then estimated as $F = 1 - p/3$, giving $F = 96.09 \pm 0.05\%$ before optimization and $F = 99.359 \pm 0.008\%$ after optimization.

\subsection{Single-qubit gate tune-up}
\noindent Finally, we optimize the single-qubit gate fidelity. From the examples presented here, single qubit gate fidelity optimization is the least demanding in terms of optimization complexity: given that we here aim to rotate just one qubit at a time, we need to find the right combination of the amplitude, frequency and the duration of the microwave driving field. Yet, the frequent demand for the gate tune-up requires that this task is performed with high speed and reliability. In this instance, we choose as a cost function the return probability $P_r$ for a randomized benchmarking (RB) sequence. The choices of the specific parameters of the RB evaluation are specific to the device and are determined beforehand via a series of optimization runs with a low number of generations and a small population size. Here, as hyperparamemeters, we determined to use $P_r$ for $m = 30$ Clifford gates and average over 15 sequence randomizations (see Appendix~\ref{app:1Q-hyperparams}) for details.

\begin{figure}
    \centering
    \includegraphics[width=\linewidth]{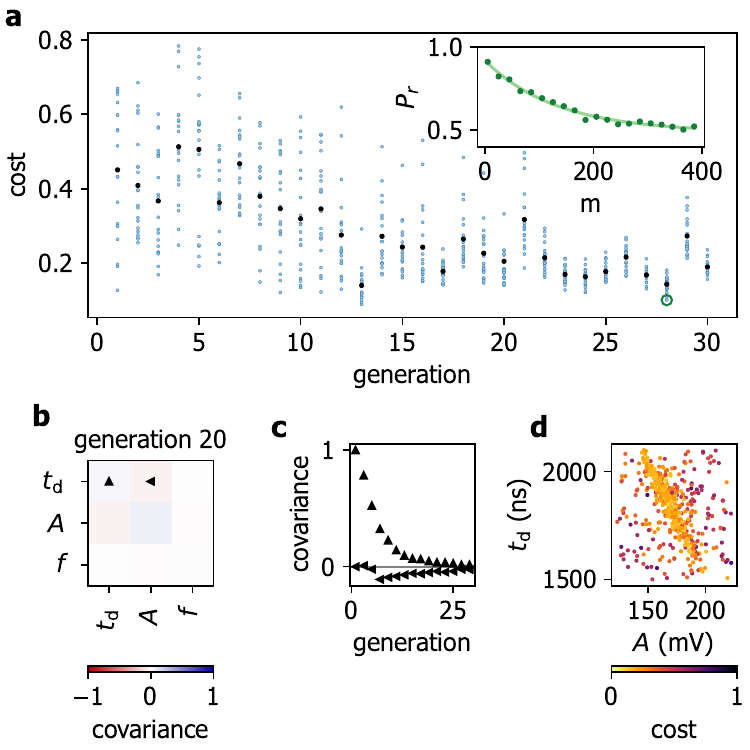}
    \caption{\textbf{Single qubit gate tune-up}.\textbf{a)} The cost function versus the generations. The best cost function value is marked (green circle). The corresponding best parameter settings are is used to run a full RB sequence (green dots) in the inset, fitted to $P_r = A p^m + C$ (light green line), yielding $p = 0.992 \pm 0.001$. The mean value per generation of the cost is plotted with black dots.  \textbf{b)} The covariance matrix at generation 20 of the driving time $t_d$, the driving amplitude $A$, and the driving frequency, $f$. \textbf{c)} The covariance matrix of element ($t_d, A$) versus the generation. \textbf{d)} A color plot of the cost versus the parameter pair $A$ and $t_d$.}
    \label{fig:single_q_gate}
\end{figure}

In Fig.~\ref{fig:single_q_gate}a we show the cost function as a function of the algorithm generation. The optimization consists of 30 generations with a population size of 20. The best cost function value, after convergence, is marked with a green circle. Notably, the algorithm found a very low cost function values in  earlier generations as well. We use the settings corresponding to the green circle configuration to execute a full RB procedure shown in the inset of Fig. \ref{fig:single_q_gate}a. By fitting the return probability function $P_r = Ap^m + C$, we obtain $p = 0.9925 \pm 0.0007$, which in turn corresponds to an average gate fidelity of $F = 99.80 \pm 0.02\%$, which is slightly higher than the best value $(99.65\%)$ we could tune manually in this device without automated tuning in a similar time-frame.

In this instance we know a priori that the driving time, $t_d$, and the amplitude, $A$, of the microwave drive should be inversely proportional to each other. As such we can immediately inspect the properties of the covariance matrix to determine whether the tuning algorithm also extracted this relationship. The snapshot of the covariance matrix in Fig.~\ref{fig:single_q_gate}b indeed shows a negative covariance in the element ($t_d, A$). Fig.~\ref{fig:single_q_gate}c then shows how this anti-correlation developed during the CMA-ES optimization, compared to the variance of the driving time. Additionally, the density plot of the cost function as a function of both amplitude $A$ and driving time $t_d$ in Fig.~\ref{fig:single_q_gate}d shows that the highest cost function values were densely sampled along a line with a negative slope. This observation highlights the strength of the CMA-ES in efficiently optimizing qubit control with dependent variables. 

\section{Discussion and Outlook}

In this work we introduced a contemporaneous, on-device global optimization as a new means to optimize the operation of spin qubits. We demonstrated that the use of the global optimizer can efficiently tackle a diverse set of semiconductor quantum processor operation tasks. While the tune-up of readout, shuttling and gate operation is still accessible for human operators on the 6-dot device we present here, all of these tasks are time-consuming, in some cases need to be repeated periodically, or have inconsistent outcomes. The automated routines presented here generally ran in similar or shorter time (2,5 hours for readout, 7 hours for shuttling, and  20 minutes for single-qubit gate tune-up) than the expert operators would spend. Looking at the cost function behavior as a function of generations one can notice that in all three cases the cost function converged much before the total duration of the run elapsed. Specifically, the shuttling cost function converged at 15\% of the total time (ca 1 hour); PSB readout reached 98\% visibility after 40\% of the total time (also ca 1 hour); and gate tune-up reached very high fidelities already in earlier generations as well. Going forward, when optimizing for speed, one can further optimize the number of generations and the number of measurements taken to further decrease the duration of the optimization. It is also interesting to explore the identified parameter dependencies through covariance matrix analysis in order to decrease the effective dimension of the optimization landscape. Overall, the identified parameter configurations presented here yield similar or higher performance compared to the best human operator tuning. For instance, the best voltage configuration for the conveyor shuttling presented in Ref.~\cite{de2024high} was achieved through a combination of this methodology and manual tuning. Specifically, the optimizer automatically resolved crosstalk issues, thus enabling manual tuning into the optimal operation point.

Generally, automated tuning of the semiconductor quantum computing platforms is seen as a top candidate for the state-of-art artificial intelligence methods due to high variability of the devices and complexity of the noise. While this is true when one approaches the tuning problem from the gradient navigation perspective, here we show that even relatively simple gradient-free algorithm can find optima in the noisy operation landscapes just as well. Without the requirements for training or labeled datasets, this approach is also compact and immediately reusable across a range of experiments. Another striking advantage is our ability to navigate a large number of parameters simultaneously without excessive runtime overhead. Regressing on more than 10 parameters simultaneously with a neural network (or other gradient based method) can be very costly \cite{valenti2022scalable}, especially in the presence of a noisy landscape. In the device, whose operation we report on here, any tuning task can be formulated as an optimization for $< 20$ parameters. This already constitutes a state-of-the-art challenge and further scaling remains to be tested on larger devices. While larger devices will have more gates and readout dots to tune-up, the optimization protocol we present here could be ran in parallel on different parts of the device. The key hurdle in scaling to simultaneous operation of the large-scale chips will be crosstalk. Specifically, it will influence the effective dimension of the optimization we need to run for each qubit. Direct analysis of covariances as we have shown it here will be crucial in determination of key variable dependencies.

\section*{Acknowledgments}

We appreciate software support from S.L. de Snoo and fabrication and design support from L. Tryputen and S. V. Amitonov. We gratefully acknowledge discussions with Joseph Rogers, Daniel Jirovec, and Pablo Cova Fari\~na. We acknowledge financial support from
the Dutch Ministry for Economic Affairs through the allowance for Topconsortia for Knowledge
and Innovation (TKI), the “Quantum Inspire–the Dutch Quantum Computer
in the Cloud” project (Project No. NWA.1292.19.194) of the NWA research
program “Research on Routes by Consortia (ORC),” which is funded by the Dutch
Research Council (NWO), the European
Union’s Horizon 2020 research and innovation programme under
grant agreement no. 951852 (QLSI project), Intel Corporation and the Army Research
Office (ARO) under grant number W911NF-17-1-0274
and W911NF2310110.The views and conclusions contained in this document are those
of the authors and should not be interpreted as representing
the official policies, either expressed or implied, of the ARO
or the US Government. The US Government is authorized
to reproduce and distribute reprints for government purposes
notwithstanding any copyright notation herein.
This research was supported by the European Union’s Horizon Europe programme under the Grant Agreement 101069515 – IGNITE. This work is also part of the project Engineered Topological Quantum Networks (Project No.VI.Veni.212.278) of the research program NWO Talent Programme Veni Science domain 2021 which is financed by the Dutch Research Council (NWO).

\section*{Author Contributions}

EG conceptualized the project with input from SRKF, YM, BU, VG and LMKV. SRKF and YM implemented the code with the input of BU, MDS, VG, KC, IFF and EG. YM, MDS, BU, KC, IFF contributed to experimental workflow. SRKF, YM, MDS, IFF and KC measured the experimental data and analyzed the results with input from BU, LMKV and EG. CVM and MRR developed the theory for parameter contribution assessment. GS provided the enabling 28Si/SiGe heterostructure. SRKF, YM, CVM, LMKV and EG co-wrote the paper with input from all authors. LMKV and EG supervised the project.

\section*{Code and Data}
The code and data to reproduce the results presented in this work can be found at this URL \url{https://gitlab.com/QMAI/papers/evolutionarytuning}.

\appendix

\section{CMA-ES}
\label{app:CMA-ES}

The algorithm underlying our optimization pipeline is called the \textit{covariance matrix adaptation evolution strategy} (CMA-ES) \cite{hansen2006cma}. The CMA-ES is a numerical optimization method that belongs to the class of evolution strategies, which are inspired by the theory of evolution in biological systems.

An evolution strategy (ES) is an iterative process used to optimize a cost function  $f(\mathbf{x}): \mathbb{R}^n \rightarrow \mathbb{R}$ with respect to its variables $\mathbf{x} \in \mathbb{R}^n$. The optimization is formulated as a minimization problem, meaning that a solution $\mathbf{x}$ is better than another solution $\mathbf{x}^\prime$ when $f(\mathbf{x}) < f(\mathbf{x}^\prime)$. In $g+1^{th}$ iteration, an ES generates new candidate solutions based on the previous, $g^{th}$, iteration. In evolution strategies, each candidate solution is referred to as an \emph{individual}, and the set of individuals that are generated in one iteration of the algorithm is called a \emph{generation} $g$.

\begin{figure}
    \centering    \includegraphics[width=\linewidth]{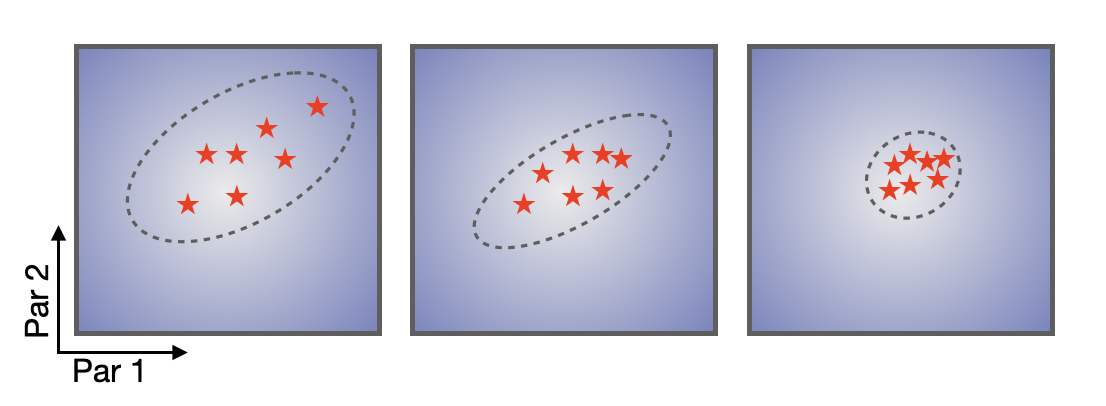}
    \caption{An illustration of the CMA-ES convergence mechanism: the purple to white transition denotes the true underlying distribution with white area corresponding to the lowest value of the cost function. The dashed ellipse shows the shape of the sampled distribution and red stars correspond to the sampled individuals. Each square corresponds to one generation of the optimization. From left to right, the optimizer updates the distribution such that more samples are drawn from the vicinity of the objective minimum.}
    \label{fig:CMA-ES}
\end{figure}

In the CMA-ES specifically, individuals are randomly sampled from a normal distribution $\mathcal{N}(\mathbf{m}, \sigma^2\mathbf{C})$, with mean vector $\mathbf{m}$, step size $\sigma$, which can be regarded as the overall spread of the distribution, and the covariance matrix $\mathbf{C}$. By adapting these three parameters for each generation, the CMA-ES should direct the distribution towards regions in the search space where good solutions are found. Figure~\ref{fig:CMA-ES} depicts several generations in an example optimization run. 

\noindent Sampling the next generation of individuals $\mathbf{x}^{(g+1)}_i$ for $i = 1, 2, \dots, \lambda$ from the normal distribution $\mathcal{N}(\mathbf{m}^{(g)}, (\sigma^{(g)})^2 \mathbf{C}^{(g)})$ is equivalent to first sampling $\mathbf{z}^{(g+1)}_i$ from a normal distribution with unit covariance matrix $\mathbf{I}$ and unit $\sigma$, and then transforming the samples in two consecutive steps:

\begin{equation} \label{eq:sampling}
\begin{aligned}
\mathbf{z}^{(g+1)}_i  &\sim \mathcal{N}(\mathbf{0}, \mathbf{I}), \\
\mathbf{y}^{(g+1)}_i  =\sqrt{\mathbf{C}^{(g)}} \mathbf{z}^{(g+1)}_i \quad &\sim \mathcal{N}(\mathbf{0}, \mathbf{C}^{(g)}), \\
\mathbf{x}^{(g+1)}_i  =\mathbf{m}^{(g)}+\sigma^{(g)} \mathbf{y}^{(g+1)}_i \quad &\sim \mathcal{N}\left(\mathbf{m}, (\sigma^{(g)})^2 \mathbf{C}^{(g)}\right).\\
\end{aligned}
\end{equation}

\noindent The vectors resulting from the first transformation $\mathbf{y}^{(g+1)}_i$ are sampled from a distribution with zero mean and an unscaled covariance matrix $\mathbf{C}$. This matrix captures the directional dependencies between control variables but does not include the overall spread, which is controlled separately by $\sigma$. In the CMA-ES, $\mathbf{C}$ and $\sigma$ are adapted separately in order to obtain competitive search speed.

In summary, each generation of the CMA-ES is comprised of the following three steps:

\begin{enumerate}
    \item Sample generation $g+1$ of $\lambda$ individuals from the current normal distribution $\mathcal{N}(\mathbf{m}^{(g)}, (\sigma^{(g)})^2 \mathbf{C}^{(g)})$.
    \item Evaluate the individuals and rank them from lowest function evaluation (best) to highest function evaluation (worst).
    \item According the ranking in step 2, update the mean $\mathbf{m}^{(g+1)}$, step size $\sigma^{(g+1)}$ and covariance matrix $\mathbf{C}^{(g+1)}$.
\end{enumerate}

A detailed tutorial on how the parameters of the normal distribution evolve is given in Ref.~\cite{hansen2006cma}, which is basis for the implementation in the Python package $\verb|cmaes|$~\cite{nomura2024cmaes} that we used in our experiments.

\begin{figure*}
    \centering
    \includegraphics[width = \linewidth]{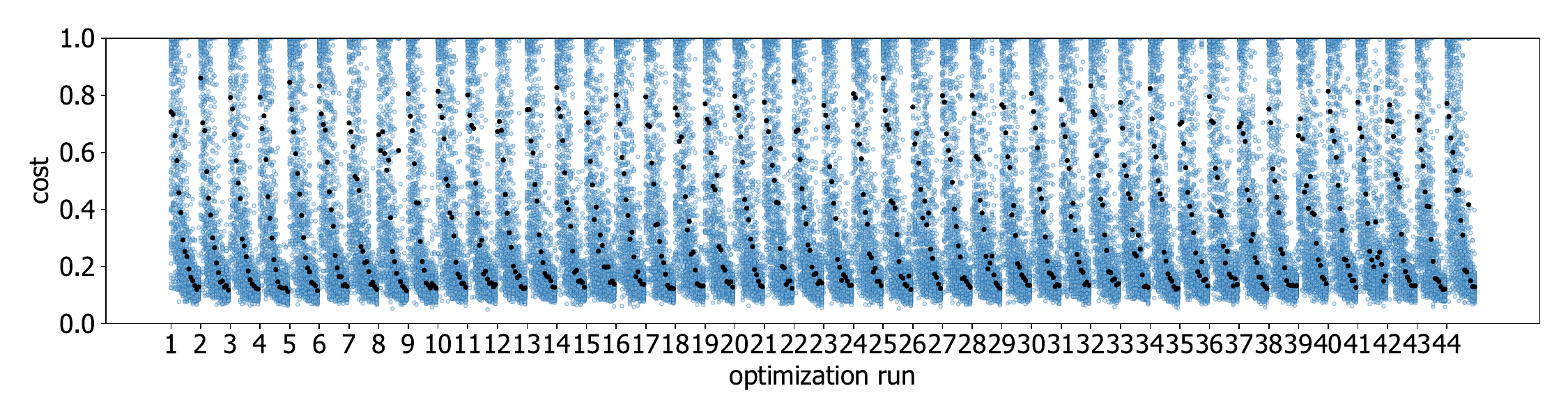}
    \caption{Repeated readout pulse optimization on secondary 6 dot device with a linear architecture. We show the cost function Eq.~\eqref{eq:visibility} evolution for 44 optimizations, each consisting of 15 generations.
    The optimized parameters are the voltage amplitudes during readout and before initialization of plungers 1 and 2, $vP_\mathrm{1,read}, vP_\mathrm{2,read}, vP_\mathrm{1,init}, vP_\mathrm{2,init}$, barrier voltages 0-2 during readout, $B_\mathrm{0,read}, B_\mathrm{1,read}, B_\mathrm{2,read}$ and barrier voltage 1 before initializaiton $B_\mathrm{1,init}$. Furthermore, the ramp times between the readout stages were added $t_\mathrm{zero,read}, t_\mathrm{read,init}, t_\mathrm{init,zero}$. The obtained cost values are plotted for each individual (blue dots) along with the generation mean (black dots).}
    \label{fig:app:readout-repeat}
\end{figure*}

We integrated the CMA-ES implementation from the package \verb|cmaes| into the existing software framework consisting of the Python packages \verb|core_tools| \cite{core_tools}, \verb|pulselib| \cite{pulse_lib}, and \verb|qconstruct|. We create and execute the circuits and subsequently evaluate the measurements to obtain a value for the cost function $f$ corresponding to the sampled individuals $\mathbf{x}$ during the CMA-ES optimization run. Moreover, we allow variables to be chosen from a settings file which is part of the existing software and is also used for manual tuning. This allows users to easily transition between manual calibrations and automated fine-tuning. 

For each optimization task, the population size and number of generations were chosen experimentally and the initial step size was always set to 1. All other hyperparameters were set to the default values of the CMA-ES implementation~\cite{nomura2024cmaes}.

\section{Readout optimization consistency}
\label{app:readout-consistency}
To test the consistency and transferability of the procedure, we performed a large scale repeated scan on another Si/SiGe device with linear architecture shown in Fig.~\ref{fig:scheme} and described in detail in Ref.~\cite{philips2022universal}. Repeating the pulse sequence outlined in Fig.~\ref{fig:psb}a. We repeated the 15 generation optimization run for 44 times. The optimized parameters are the voltage amplitudes during readout and before initialization of plungers 1 and 2, $vP_\mathrm{1,read}, vP_\mathrm{2,read}, vP_\mathrm{1,init}, vP_\mathrm{2,init}$, barrier voltages 0-2 during readout, $B_\mathrm{0,read}, B_\mathrm{1,read}, B_\mathrm{2,read}$ and barrier voltage 1 before initializaiton $B_\mathrm{1,init}$. Furthermore, the ramp times between the readout stages were added $t_\mathrm{zero,read}, t_\mathrm{read,init}, t_\mathrm{init,zero}$. The results of the optimization are shown in Fig.~\ref{fig:app:readout-repeat}: all the 44 runs have converged and the visibilities between 85\% and 90\% (fidelities between 92.5\% and 95\%) were consistently reached.

\section{Cost function hyperparameters of single-qubit gate optimization}
\label{app:1Q-hyperparams}
\begin{figure}[b]
    \centering
    \includegraphics[width=\linewidth]{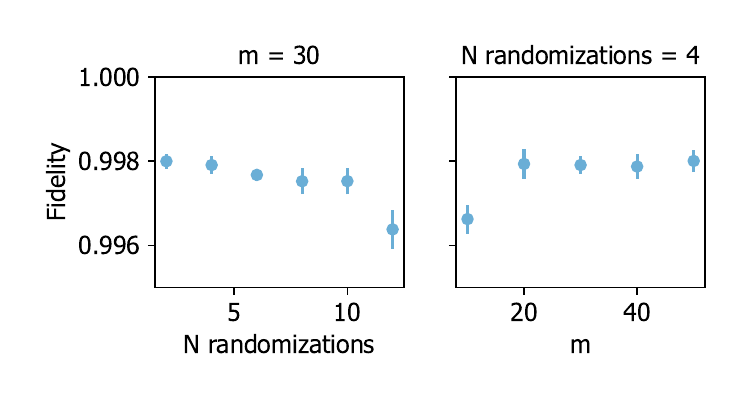}
    \caption{The single qubit gate fidelity obtained after the optimization by using different cost function hyperparameters. The left panel shows the fidelity against the number of randomizations in the cost function for a fixed sequence length of $m = 30$. The right panel shows the fidelity against a varying sequence length for a fixed number of randomizations.}
    \label{fig:app:SQG_hyper}
\end{figure}

The cost function we used for the single-qubit gate optimization shown in Fig. \ref{fig:single_q_gate} used a sequence length, $m$, of 30 Clifford gates, which was averaged over four sequence 
randomizations. In Fig \ref{fig:app:SQG_hyper}, we show the obtained fidelities for different sequence lengths and number of randomizations used in the cost function. We obtain fidelities  $> 99.5$\% for all cost function settings. Note that the varied hyperparameters pertain only to the cost function during optimization and not to the RB sequence from which we obtained the fidelities.

\section{Initialization pulse: Theory analysis}
\label{app:init-theory}

\subsection{Theoretical model}

In the main text, we performed an HDMR analysis of the parameter contribution and found that the initialization pulse parameters played a crucial role once targeting high-fidelity readout. To understand this behavior, we provide a minimal model that captures the observed sensitivity to initialization. We model the double quantum dot using a generalized Hubbard model~\cite{zwolak2023colloquium}
\begin{align}
    &\hat{H}_\text{DQD} =-t_c\sum_{ij,\sigma}\Big(\hat{c}_{i,\sigma}^\dagger \hat{c}_{j,\sigma} +\text{h.c.}\Big) +\\ &\sum_{\langle ij\rangle}U_{ij}\hat{n}_i\hat{n}_j+\sum_j \left(\frac{U}{2}\hat{n}_j(\hat{n}_j-1)+\mu_j\hat{n}_j+\frac{1}{2}\mu_BB^j \sigma_z^j\right). \nonumber
\end{align}
Here $\hat{c}^\dagger_{j,\sigma}(\hat{c}_{j,\sigma})$ is the second quantization creation (annihilation) operator for an electron on site $j=L,R$ with spin $\sigma$, $\hat{n}_j=\sum_\sigma \hat{c}^\dagger_{\sigma,j}\hat{c}_{\sigma,j}$ is the number operator, and $\sigma_z^j=\hat{c}^\dagger_{\uparrow,j}\hat{c}_{\uparrow,j}-\hat{c}^\dagger_{\downarrow,j}\hat{c}_{\downarrow,j}$ is the component of the Spin operator parallel to the magnetic field. The parameters $U$ and $U_{ij}$ describe the intra- and inter-dot Coulomb repulsion, $t_c$ is the tunnel coupling, $V_i$ are the chemical potentials in each dot, and $\mu_B$ is Bohr's magneton. In the linearized Thomas Fermi approximation~\cite{zwolak2023colloquium}, the chemical potential is given by the electrostatic field generated by the gate electrodes, $\mu_i=\sum_j \alpha_{ij}V_j$, where $\alpha_{ij}$ is the virtual gate matrix and $V_j$ are the applied voltages on gate $j$.

We now also introduce the notation $(n_L, n_R)$ to describe the number of electrons in the left and right dots.
Note that while we experimentally investigated the transition between the charge regimes (3,1) and (4,0), in practice only the outer-two electrons contribute to the dynamics. This is the frozen core approximation, which allows us to ignore the low-lying orbital valley states. In this approximation, the dynamics are still described by the Hamiltonian above using two electrons in the (1,1) and (2,0) charge configuration. Ignoring excited orbital states, the following states span the relevant Hilbert space 
\begin{align}
    \ket{S(2,0)} &=\hat{c}^\dagger_{L,\uparrow}\hat{c}^\dagger_{L,\downarrow} \ket{\text{vac}} \\
    \ket{S(1,1)} &=\frac{1}{\sqrt{2}}\left(\hat{c}^\dagger_{L,\uparrow}\hat{c}^\dagger_{R,\downarrow}- \hat{c}^\dagger_{L,\downarrow}\hat{c}^\dagger_{R,\uparrow}\right)\ket{\text{vac}} \\
    \ket{T_0(1,1)} &=\frac{1}{\sqrt{2}}\left(\hat{c}^\dagger_{L,\uparrow}\hat{c}^\dagger_{R,\downarrow}+ \hat{c}^\dagger_{L,\downarrow}\hat{c}^\dagger_{R,\uparrow} \right)\ket{\text{vac}}, 
\end{align}
where $\ket{\text{vac}}$ represents the vacuum state. Note, that we have neglected the polarized triplet states $\ket{T_+}=\hat{c}^\dagger_{L,\uparrow}\hat{c}^\dagger_{R,\uparrow} \ket{\text{vac}}$ and $\ket{T_-}=\hat{c}^\dagger_{L,\downarrow}\hat{c}^\dagger_{R,\downarrow} \ket{\text{vac}}$ as their presence is decoupled for the weak spin-orbit coupling in silicon and stray magnetic fields from the micromagnet. In this basis, the Hamiltonian reads
\begin{align}
    \label{eqn: 3x3 dqd Hamiltonian}
    \hat{H}_\text{DQD,eff}=\begin{pmatrix}
        -\varepsilon(t) & t_c & 0 \\
        t_c & 0 & \Delta E_Z \\
        0 & \Delta E_Z& 0
    \end{pmatrix},
\end{align}
where we introduced the Zeeman splitting difference $\Delta E_z=\mu_B(B_R-B_L)$ and the detuning $\varepsilon=\mu_L-\mu_R$.

\begin{figure}
    \centering
    \includegraphics[width=0.8\linewidth]{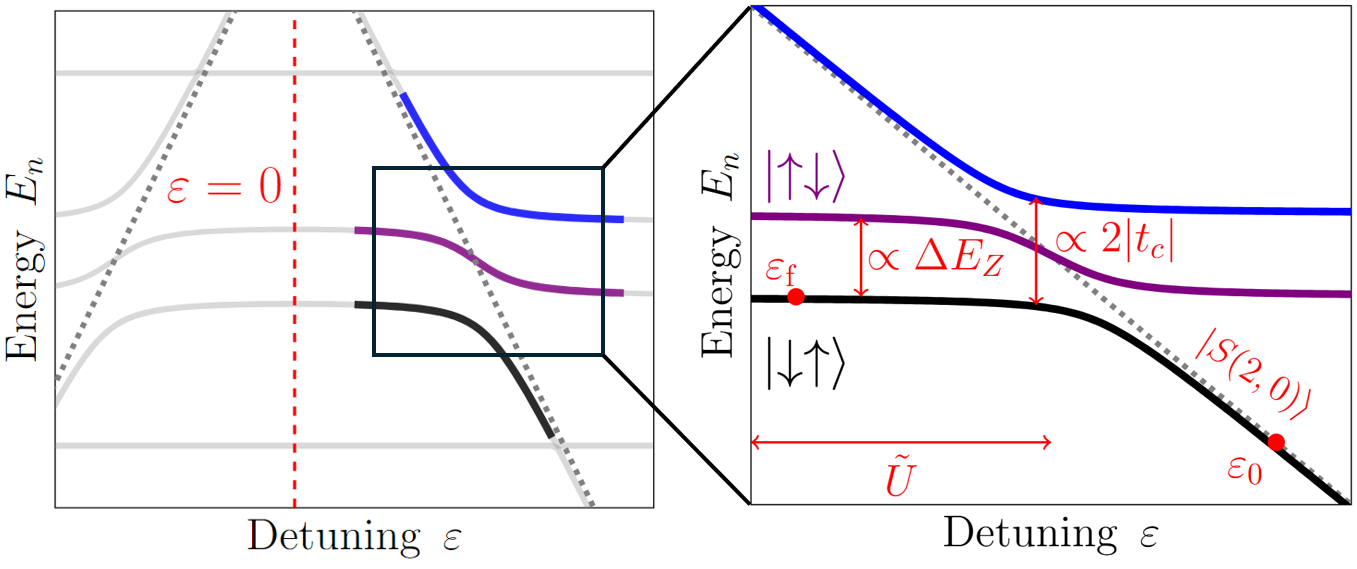}
    \caption{Energy levels of the double quantum dot model including the polarized states (left). Our effective model in Eq.~\eqref{eqn: 3x3 dqd Hamiltonian} and how it is embedded into the larger energy landscape (right). Note that in the model, we absorbed the term $\Tilde{U}=U-U_{12}$ into the detuning as it only constitutes a constant shift.}
    \label{fig: energy 3x3 dqd}
\end{figure}

Importantly, our model reveals a small anti-crossing for small values of $t_c/\Delta E_Z$, which can be seen in Figure~\ref{fig: energy 3x3 dqd}. Hence, the initialization fidelity will, in part, strongly depend on the ratio of the tunnel coupling and the Zeeman splitting difference.

\subsection{Simulation of the initialization error in PSB visibility}

We now numerically simulate the expected transfer errors as functions of varying the representative physical parameters to understand the susceptibility of the PSB visibility to initialization and readout pulse~\cite{seedhousePauliBlockadeSilicon2021}. For simplicity, we focus here on the initialization process. We model the linear ramp from the main text as
\begin{align}
    \varepsilon(t)=\Big(\varepsilon_\text{f}-\varepsilon_0\Big)\frac{t}{t_\text{f}}+\varepsilon_0.
\end{align}
where $\varepsilon_0$, is the initial detuning setting, $\varepsilon_\text{f}$ is the final detuning setting, $t_\text{f}$ is the final time, and $t=0$ is the initial time. We take into account the effect of decoherence caused by charge noise fluctuations coupling via detuning fluctuations $\varepsilon_0\rightarrow\varepsilon_0+\delta\varepsilon(t)$. Due to the short pulse times, we use the quasistatic approximation $\delta\varepsilon(t)=\delta\varepsilon$, apply a Monte-Carlo sampling method, where we draw $\delta\varepsilon$ from a Gaussian distribution with mean $\langle\delta\varepsilon\rangle=0$ and standard deviation $\sqrt{\langle\delta\varepsilon^2\rangle}=1$ GHz, and neglect high-frequency noise~\cite{macquarrieProgressCapacitivelyMediated2020}. 

\begin{figure*}
    \centering
    \includegraphics[width=.8\linewidth]{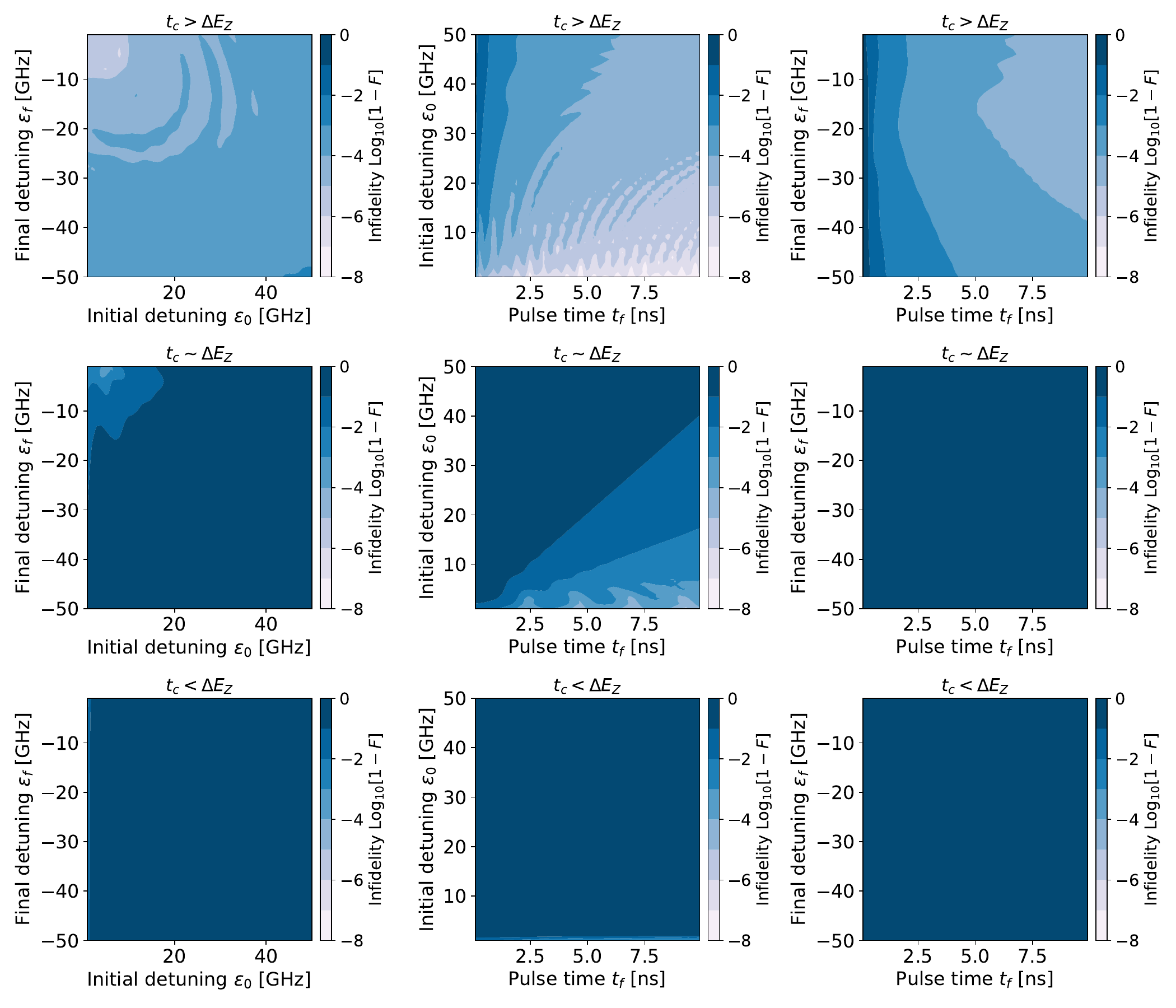}
    \caption{Simulated fidelity for spin initialization as a function of (first column: $\varepsilon_0,\varepsilon_\text{f}$), (second column: $t_\text{f},\varepsilon_0$), (third column: $t_\text{f},\varepsilon_\text{f}$) for different tunnel coupling regimes: strong ($t_c=10\text{ GHz}> \Delta E_Z=0.3$ GHz, first row), medium ($t_c=\Delta E_Z=2$ GHz, second row), and weak ($t_c=0.3\text{ GHz}< \Delta E_Z=2$ GHz, third row). Parameters used whenever they are not sweeped over: $t_\text{f}=4 \text{ GHz}, \varepsilon_0=0, \varepsilon_\text{f}=50$ GHz.}
    \label{fig: pca full}
\end{figure*}

\begin{figure*}
    \centering
    \includegraphics[width=.8\linewidth]{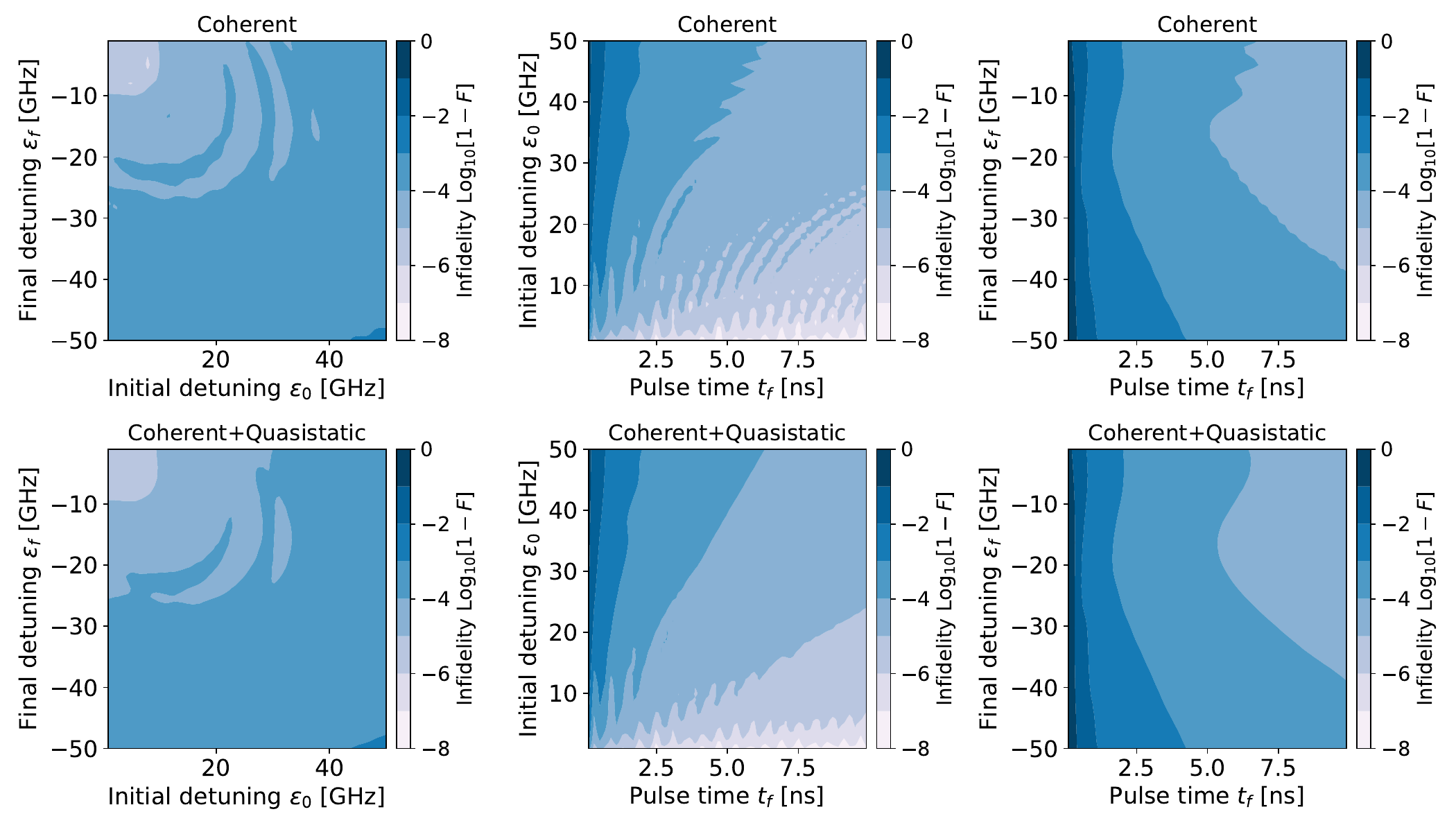}
    \caption{Simulated fidelity for spin initialization as a function of ($\varepsilon_0,\varepsilon_\text{f}$), ($t_\text{f},\varepsilon_0$), ($t_\text{f},\varepsilon_\text{f}$) in the strong tunneling regime $t_c=10$ GHz and $\Delta E_Z=0.3$ GHz without (first row) and with quasistatic noise (second row). We use the same parameter settings as in Fig.~\ref{fig: pca full} for the coherent evolution. For the incoherent dynamics, we use 1000 Monte-Carlo samples assuming quasistatic Gaussian charge noise $\delta\varepsilon(t)=\delta\varepsilon$ with $\langle\delta\varepsilon\rangle=0$ and standard deviation $\sqrt{\langle\delta\varepsilon^2\rangle}=1$ GHz.}
    \label{fig: qsn high tc}
\end{figure*}

Below, we will provide an analysis of these parameters for the single linear pulse under coherent evolution with and without quasistatic noise in the detuning. 
Our simulations in Figure~\ref{fig: pca full} reveal that the resulting fidelity depends on the following five parameters
\begin{itemize}
    \item Initial detuning point $\varepsilon_0$.
    \item Final detuning point $\varepsilon_\text{f}$.
    \item Pulse times $t_\text{f}$.
    \item Tunnel coupling $t_c$.
    \item Zeeman splitting difference $\Delta E_Z$.
\end{itemize}
Firstly, in Figure~\ref{fig: pca full}, we show the simulation results for three coupling regimes: strong ($t_c\gg \Delta E_Z$, upper row), medium ($t_c\simeq \Delta E_Z$, middle row), and weak ($t_c\ll \Delta E_Z$, bottom row) as a function of $\varepsilon_0$ and $\varepsilon_\text{f}$, $t_f$ and $\varepsilon_0$, and $t_f$ and $\varepsilon_\text{f}$. We see that these parameters can alter the behavior of the charge transfer probability due to Landau-Zener transitions~\cite{ivakhnenkoNonadiabaticLandauZener2023}. For instance, in our experiment, we find that the PSB visibility is enhanced for QD5-QD6 and QD1-QD2. This could be explained by the different Zeeman splitting differences of QD5-QD6 and QD1-QD2, thus, different $t_c/\Delta E_Z$. In Figure~\ref{fig: qsn high tc}, we show the results, including quasistatic charge noise. While quasistatic noise reduces the overall fidelity, the general trend is still similar to the coherent dynamics. Importantly, high-fidelity readout $>99\%$ is only expected in very specific regions in parameter space. This emphasizes the importance of precise pulse control for optimizing PSB visibility~\cite{meinersenQuantumGeometricProtocols2024}.

\section{\label{supp:subsec:twotone} Two tone conveyor pulse}
The Conveyor voltage pulse used in this experiment is shown here. In alignment with prior research \cite{de2024high}, we employ a combination of two sinusoidal waves for each gate: one operating at frequency $f$ and another at  $f/2$. This can be mathematically formulated as 

\begin{equation}
V_n(t) = V^{DC}_n + \frac{A}{2}[\sin(2\pi ft - \phi_n) + \sin(\pi ft - \theta_n)],
\end{equation}

where $V^{DC}_n$ indicates the individual gate offset, A signifies the amplitude, and $\phi_n$ and $\theta_n$ represent the phase shifts for the respective frequency components. Here, the index $n$ corresponds to the gate electrodes arranged sequentially to the right of $vP_2$. This technique produces expanded potential barriers between consecutive conveyor minima, markedly reducing the probability of charge escape during transport operations.

\bibliographystyle{unsrt}
\bibliography{evolutionary}

\newpage

\end{document}